\documentclass{webofc}
\usepackage[varg]{txfonts}   % Web of Conferences font

\usepackage{multirow}

\newcommand{\fref}[1]{Fig. \ref{#1}}
\newcommand{\tref}[1]{Tab.~\ref{#1}}

\begin{document}

\title{Glueballs from bound state equations}

\author{\firstname{Markus Q.} \lastname{Huber}\inst{1}\fnsep\thanks{\email{markus.huber@physik.jlug.de}} \and
        \firstname{Christian S.} \lastname{Fischer}\inst{1,2}\fnsep\thanks{\email{christian.fischer@theo.physik.uni-giessen.de}} \and
        \firstname{H\`elios} \lastname{Sanchis-Alepuz}\inst{3}\fnsep\thanks{\email{helios.sanchis-alepuz@silicon-austria.com}}
}

\institute{
Institut f\"ur Theoretische Physik, Justus-Liebig-Universit\"at Giessen, Heinrich-Buff-Ring 16, 35392 Giessen, Germany
\and
Helmholtz Forschungsakademie Hessen f\"ur FAIR (HFHF), GSI Helmholtzzentrum f\"ur Schwerionenforschung, Campus Giessen, 35392 Giessen, Germany
\and
Silicon Austria Labs GmbH, Inffeldgasse 33, 8010 Graz, Austria
}

\abstract{
Glueballs are bound states in the spectrum of quantum chromodynamics which consist only of gluons.
They belong to the group of exotic hadrons which are widely studied experimentally and theoretically.
We summarize how to calculate glueballs in a functional framework and discuss results for pure Yang-Mills theory.
Our setup is totally self-contained with the scale being the only external input.
We enumerate a range of tests that provide evidence of the stability of the results.
This illustrates the potential of functional equations as a continuum first-principles method complementary to lattice calculations.
}

\maketitle

\section{Introduction}
\label{sec:introduction}

The existence of glueballs was conjectured already a long time ago \cite{Fritzsch:1972jv}.
Nevertheless they remained elusive states, see, e.g., \cite{Klempt:2007cp,Crede:2008vw,Mathieu:2008me,Ochs:2013gi,Llanes-Estrada:2021evz}, in part because of the fact that they can mix with quark states.
In particular for the lightest glueball, which has scalar quantum numbers $J^{\mathsf{PC}}=0^{++}$, the existence of several states in the relevant mass range complicates any analysis.
Several scenarios and candidates were discussed in the past, most of which expect the lightest glueball to be in the range between 1500 and 2000~MeV.
Two recent analyses of radiative $J/\psi$ decay data from BESIII find evidence for a scalar glueball state between 1700 and 1900~MeV \cite{Sarantsev:2021ein,Rodas:2021tyb}.
Central exclusive production is another promising process to look for glueballs.
Corresponding data from CMS does not have high enough statistics in the relevant mass range \cite{CMS:2020jbb}, though.
Another example is $p\bar{p}$ annihilation, which is the relevant process for the future PANDA experiment \cite{PANDA:2021ozp}.

Theoretical investigations of glueballs complement the experimental searches but they are challenging as well.
Employed approaches include Hamiltonian many body methods \cite{Szczepaniak:1995cw,Szczepaniak:2003mr}, chiral Lagrangians \cite{Janowski:2011gt,Eshraim:2012jv}, lattice methods \cite{Morningstar:1999rf,Chen:2005mg,Gregory:2012hu,Brett:2019tzr,Athenodorou:2020ani,Chen:2021dvn} and functional methods \cite{Huber:2020ngt,Huber:2021yfy}.
The last two methods have provided results from first-principles in the case when quarks are neglected.
In that case, only pure gluonic states exist what alleviates the analysis.
There is some progress in the inclusion of quarks on the lattice, but no final conclusions as to how large this unquenching effect is could be drawn yet, see, e.g., \cite{Gregory:2012hu,Brett:2019tzr,Chen:2021dvn}.

In this contribution we describe the status of glueball calculations with functional methods.
Using bound state equations, the glueball spectrum is calculated for spin $J=0,1,2,3,4$ for pure Yang-Mills theory.
As input, the correlation functions of the elementary gluon and ghost fields in Landau gauge are required.
In contrast to mesons, baryons and even tetraquarks, see, e.g., \cite{Cloet:2013jya,Eichmann:2016yit,Eichmann:2020oqt}, literature on functional glueball calculations is scarce and was restricted to scalar and pseudoscalar glueballs \cite{Meyers:2012ka,Sanchis-Alepuz:2015hma,Souza:2019ylx,Kaptari:2020qlt} until recently \cite{Huber:2020ngt,Huber:2021yfy}.
The difference between pure quark bound states and states including gluons is that for the former simpler approximations for the input are often sufficient \cite{Cloet:2013jya,Eichmann:2016yit,Eichmann:2020oqt}.
Calculations for glueballs, on the other hand, depend very sensitively on the input.
For example, using simple models for the vertices, it is possible to obtain a decent result for the scalar glueball mass, but using the same model does not describe the pseudoscalar glueball well.
The progress made in the last decade in the calculation of correlation functions, see \cite{Huber:2018ned} and references therein, has led to quantitatively reliable results for various correlation functions \cite{Cyrol:2016tym,Huber:2020keu}.
The availability of such high quality input was decisive to obtain the results presented below.

We shortly describe the setup of the calculations in the next section.
In Sec.~\ref{sec:results} we present the results and in Sec.~\ref{sec:stability} we discuss the stability of the results with regard to the employed approximations.
We conclude in Sec.~\ref{sec:conclusions}.

\section{Bound state equations}
\label{sec:bses}

% BSE
\begin{figure}
 \begin{center}
  \includegraphics[width=0.8\textwidth]{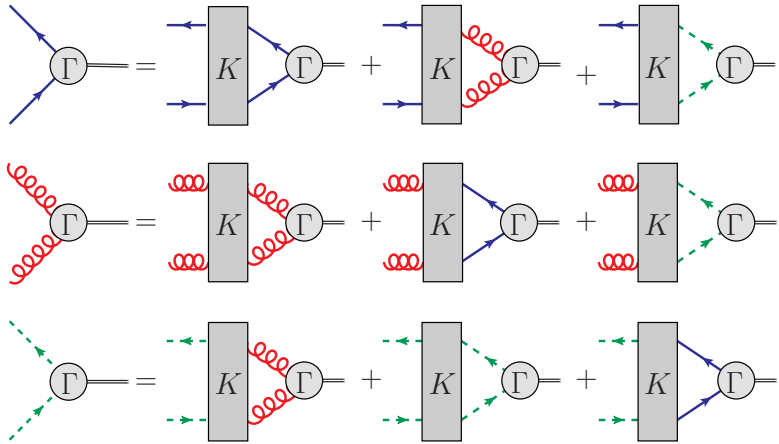}
 \end{center}
 \caption{
    Two-body bound state equations of QCD.
    The blue continuous/red wiggly/green dashed lines are quark/gluon/ghost propagators.
    $\Gamma$ are Bethe-Salpeter amplitudes and $K$ the scattering kernels.
    For pure Yang-Mills theory, diagrams with quarks do not exist.
    }
 \label{fig:bses}
\end{figure}

We calculate glueballs from two-body bound state equations (BSEs).
The full system for QCD is shown in \fref{fig:bses}.
For pure Yang-Mills theory, the diagrams containing quark propagators are neglected.

To solve the BSEs, the propagators and scattering kernels $K$ are required as input.
The latter are derived from the 3PI effective action truncated at three loops \cite{Berges:2004pu,Sanchis-Alepuz:2015tha}.
This leads to the expressions discussed in Sec.~\ref{sec:truncation} and shown in \fref{fig:kernels}.
The quantities needed to calculate them are the gluon and ghost propagators and the three-gluon and ghost-gluon vertices.
They are calculated from their equations of motion derived from the same action.
The input is discussed in more detail in Sec.~\ref{sec:input} where also comparisons with lattice results are shown in Figs.~\ref{fig:props} and~\ref{fig:tg_comp} .

The BSEs are solved as eigenvalue equations with the total momentum $P$ of the bound state as parameter.
A solution is found when an eigenvalue is one by varying $P^2$.
The corresponding mass is given by $M=\sqrt{-P^2}$.
This entails that the internal quantities need to be known for complex arguments.
However, calculations in the complex plane have not progressed as far as for Euclidean momenta, see, e.g., \cite{Fischer:2020xnb,Horak:2022myj}.
Thus we resort to calculations for $P^2>0$ and analytically continue the eigenvalue curve via Schlessinger's continued fraction method \cite{Schlessinger:1968spm,Tripolt:2018xeo}, for details see Ref.~\cite{Huber:2020ngt}.
We tested this method with a meson example where the full calculation for $P^2<0$ can be done \cite{Huber:2020ngt}.

\section{Results}
\label{sec:results}

% spectrum
\begin{figure*}[tb]
	\begin{center}
	\includegraphics[width=0.63\textwidth]{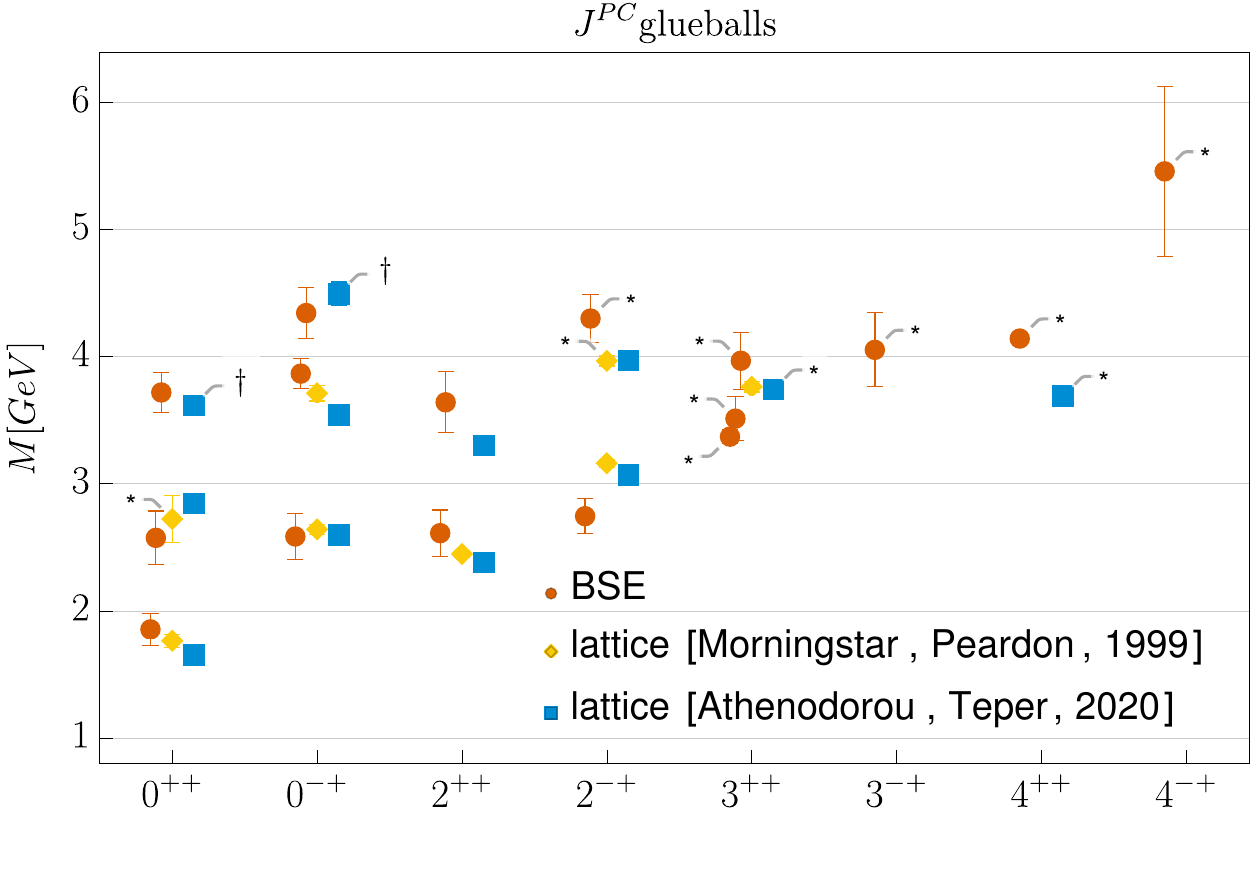}\\
	\includegraphics[width=0.63\textwidth]{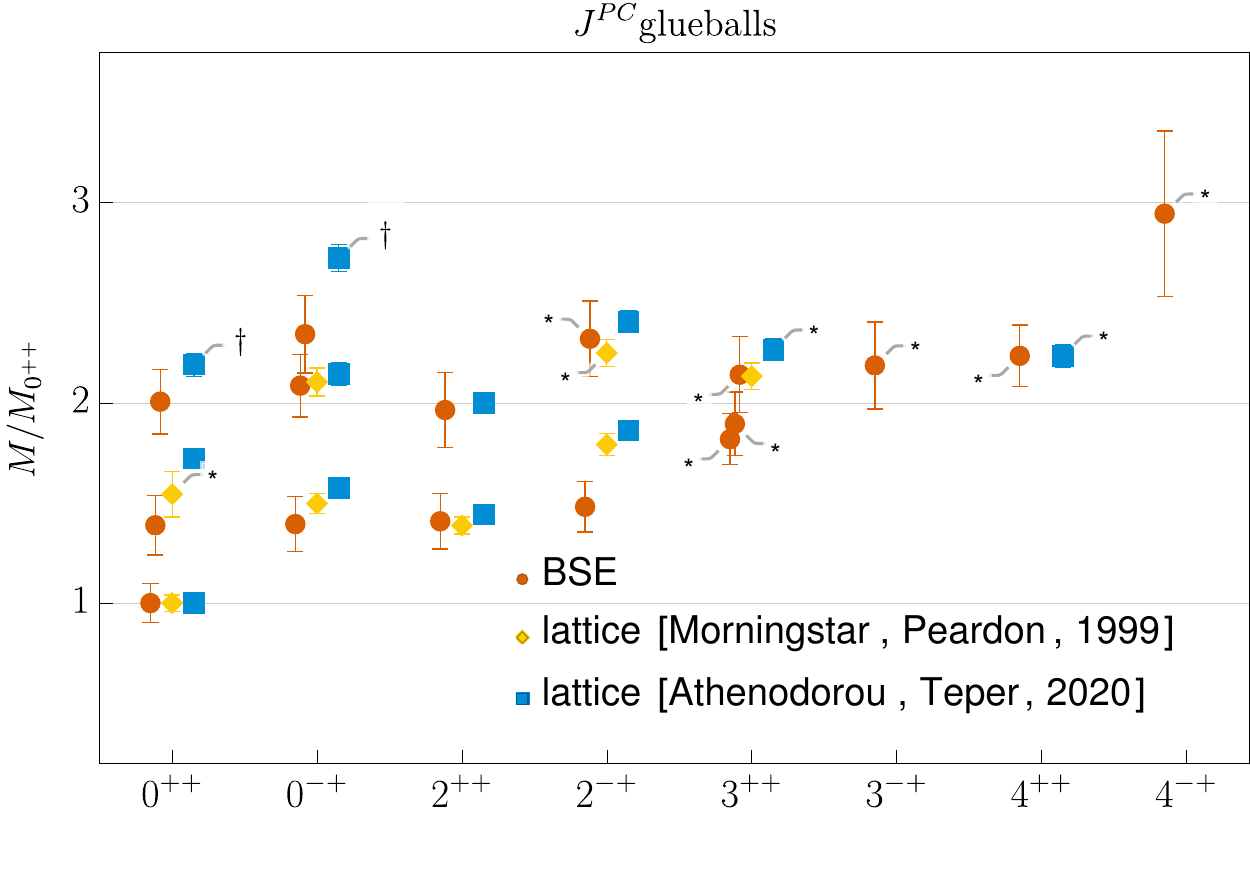}
	\end{center}
	\caption{
		Results for glueball ground states and excited states for the indicated quantum numbers from lattice simulations \cite{Morningstar:1999rf,Athenodorou:2020ani} and functional equations.
		In the upper plot, we display the glueball masses on an absolute scale set by $r_0=1/418(5)\,\text{MeV})$.
	    In the lower plot, we display the spectrum relative to the ground state.
	    Masses with $^\dagger$ are conjectured to be the second excited states.
		Masses with $^*$ come with some uncertainty in their identification in the lattice case or in the trustworthiness of the extrapolated value in the BSE case.
		}
	\label{fig:spectrum}
\end{figure*}

% spectrum in numbers
\begin{table*}[tb]
	\begin{center}
\caption{Ground and excited state masses $M$ of glueballs for various quantum numbers.
		Compared are lattice results from \cite{Morningstar:1999rf,Athenodorou:2020ani} with the functional results of  \cite{Huber:2020ngt,Huber:2021yfy}.
        For \cite{Morningstar:1999rf}, the errors are the combined errors from statistics and the use of an anisotropic lattices.
		For \cite{Athenodorou:2020ani}, the error is statistical only.
		In our results, the error comes from the extrapolation method and should be considered a lower bound on errors.
		All results use the same value for $r_0=1/(418(5)\,\text{MeV})$.
		The related error is not included in the table.
		Masses with $^\dagger$ are conjectured to be the second excited states.
		Masses with $^*$ come with some uncertainty in their identification in the lattice case or in the trustworthiness of the extrapolated value in the BSE case.
		}
		\begin{tabular}{|l||c|c|c|c|c|c|}
			\hline
			&  \multicolumn{2}{c|}{\cite{Morningstar:1999rf}} & \multicolumn{2}{c|}{\cite{Athenodorou:2020ani}} & \multicolumn{2}{c|}{This work}\\   
			\hline
			State &  $M\, [\text{MeV}]$& $M/M_{0^{++}}$ & $M\, [\text{MeV}]$& $M/M_{0^{++}}$ & $M\,[\text{MeV}]$ & $M/M_{0^{++}}$\\   
			\hline\hline
			$0^{++}$ & $1760 (50)$ & $1(0.04)$ & $1651(23)$ & $1(0.02)$ & $1850 (130)$ & $1(0.1)$\\
			\hline
			$0^{^*++}$ & $2720 (180)^*$ & $1.54(0.11)^*$ & $2840(40)$ & $1.72(0.034)$ & $2570 (210)$ & $1.39(0.15)$\\
			\hline
			\multirow{2}{*}{$0^{^{**}++}$} & \multirow{2}{*}{--} & \multirow{2}{*}{--} & $3650(60)^\dagger$ & $2.21(0.05)^\dagger$ & \multirow{2}{*}{$3720 (160)$} & \multirow{2}{*}{$2.01(0.16)$}\\
			& & & $3580(150)^\dagger$ & $2.17(0.1)^\dagger$ & &\\
			\hline
			$0^{-+}$ & $2640 (40) $ & $1.50(0.05)$ & $2600(40)$ & $1.574(0.032)$ & $2580 (180)$ & $1.39(0.14)$\\
			\hline
			$0^{^*-+}$ & $3710 (60)$ & $2.10(0.07)$ & $3540(80)$ & $2.14(0.06)$ & $3870 (120)$ & $2.09(0.16)$\\
			\hline
			\multirow{2}{*}{$0^{^{**}-+}$} & \multirow{2}{*}{--} & \multirow{2}{*}{--} & $4450(140)^\dagger$ & $2.7(0.09)^\dagger$ & \multirow{2}{*}{$4340 (200)$} & \multirow{2}{*}{$2.34(0.19)$}\\
			& & & $4540(120)^\dagger$ & $2.75(0.08)^\dagger$ & &\\
			\hline
			\hline
			$2^{++}$ & 2447(25) & 1.39(0.04) & 2376(32) & 1.439(0.028) & 2610(180) & 1.41(0.14)\\
			\hline
			$2^{^*++}$ & -- & -- & 3300(50) & 2(0.04) & 3640(240) & 1.96(0.19)\\
			\hline
			$2^{-+}$ & 3160(31) & 1.79(0.05) & 3070(60) & 1.86(0.04) & 2740(140) & 1.48(0.13)\\
			\hline
			$2^{^*-+}$ &  3970(40)$^*$ & 2.25(0.07)$^*$ & 3970(70) & 2.4(0.05) & 4300(190) & 2.32(0.19)\\
			\hline\hline
			$3^{++}$ & 3760(40) & 2.13(0.07) & 3740(70)$^*$ & 2.27(0.05)$^*$ & 3370(50)$^*$ & 1.82(0.13)$^*$\\
			\hline
			$3^{^*++}$ & -- & -- & -- & -- & 3510(170)$^*$ & 1.89(0.16)$^*$\\
			\hline
			$3^{^{**}++}$ & -- & -- & -- && 3970(220)$^*$ & 2.14(0.19)$^*$\\
			\hline
			$3^{-+}$ & -- & -- & -- & -- &  4050(290)$^*$ & 2.19(0.22)$^*$\\
			\hline
			\hline
			$4^{++}$ & -- & -- & 3690(80)$^*$ & 2.24(0.06)$^*$ & 4140(30)$^*$ & 2.23(0.15)$^*$\\
			\hline
			$4^{-+}$ & -- & -- & -- & -- & 5050(700)$^*$ & 2.9(0.4)$^*$\\
			\hline
		\end{tabular}
		\label{tab:masses}
	\end{center}
\end{table*}

The masses of the ground state and up to two excited states were calculated for spin $J=0,1,2,3,4$ and positive charge parity.
For $J=1$, no results were found which means that such states, if they exist, are very heavy.
Note that the Landau theorem \cite{Landau:1948kw,Yang:1950rg} does not apply to our framework because the gluons are not on-shell.
All found states are listed in \tref{tab:masses} and shown in \fref{fig:spectrum} in comparison to lattice results.
To have a common scale for physical units, we set the Sommer scale $r_0$ to the same value for all sources.
Alternatively, we show the spectrum in units of the lightest state.

The results for scalar, pseudoscalar and tensor glueballs agree very well with lattice results.
For the $2^{-+}$, our ground state is lower than lattice results and for $3^{++}$ we find three states rather close to each other.
This could be an artifact of the extrapolation of the eigenvalue curves which might merge to only one state at the physical mass.
In general, the extrapolation becomes less reliable for high masses which can be seen by the increase in the error bars.
The $4^{++}$ state is an exception, as its extrapolation is remarkably stable.
The states $3^{-+}$ and $4^{-+}$ were not observed on the lattice.
We would like to add that the results for the latter has changed compared to \cite{Huber:2021yfy} because more points for the extrapolation became available from the computationally very expensive calculation.
Overall, the agreement with lattice is rather good.

\section{Stability}
\label{sec:stability}

Beyond the good agreement with lattice results, alternative tests are of course advantageous.
Here we discuss two possibilities.
The first concerns the employed input, the second the bound state equations themselves.

\subsection{Input}
\label{sec:input}

The input in form of the gluon and ghost propagators and the ghost-gluon and three-gluon vertices was calculated from the 3PI effective action truncated at three-loop order \cite{Huber:2020keu}.
Solving for all these quantities at the same time does not leave any quantities to be modeled or tuned.
Consequently, the resulting system of equations is self-contained.
This is actually rather restrictive, because it removes any freedom of tuning to improve the results.
For example, it is possible to achieve good agreement of the propagators with lattice results when an effective three-gluon vertex model is used \cite{Huber:2012kd}.
However, since such a model is optimized for the gluon propagator, other equations like BSEs can be affected negatively.
Thus, such a model is not adequate for use in a BSE.

% gh, gl, 3g
\begin{figure*}[t]
	\includegraphics[width=0.48\textwidth]{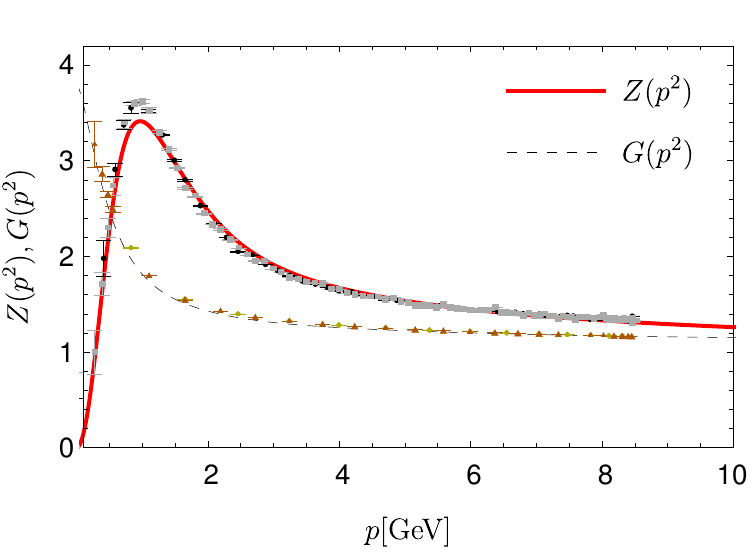}\hfill
	\includegraphics[width=0.48\textwidth]{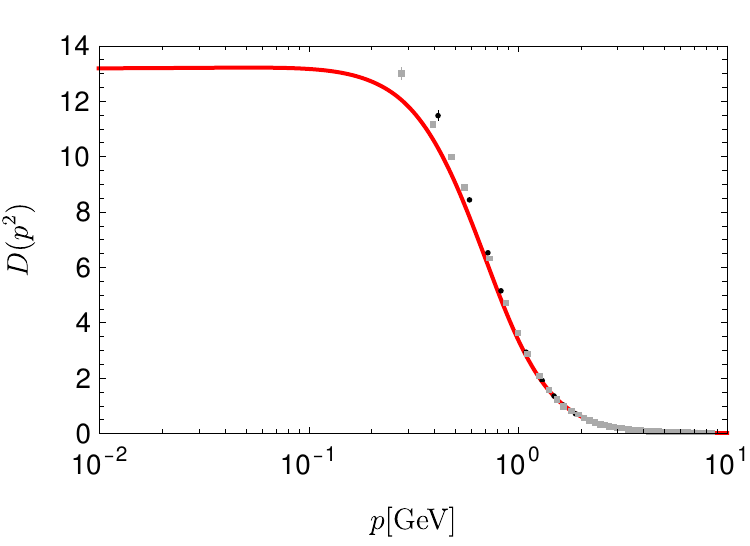}
	\caption{Gluon and ghost dressing functions $Z(p^2)$ and $G(p^2)$, respectively, (left) and gluon propagator $D(p^2)$ (right) \cite{Huber:2020keu} in comparison to lattice data \cite{Sternbeck:2006rd}.
	}
	\label{fig:props}
% ghg, 3g
% 	\includegraphics[width=0.48\textwidth]{pGhgFullSystemLattAM}\hfill
% 	    \includegraphics[width=0.48\textwidth]{pTgFullSystemSPLattAMAS}
% 	\caption{Left: Ghost-gluon vertex dressing function (full kinematic dependence) \cite{Huber:2020keu} in comparison to $SU(2)$ lattice data \cite{Maas:2019ggf}.
% 	Right: Three-gluon vertex dressing function at the symmetric point \cite{Huber:2020keu} in comparison to lattice data \cite{Cucchieri:2008qm,Sternbeck:2017ntv}).
% 	}
%     \label{fig:verts}
\end{figure*}

% 3g: comparison
\begin{figure*}[t]
	\includegraphics[width=0.48\textwidth]{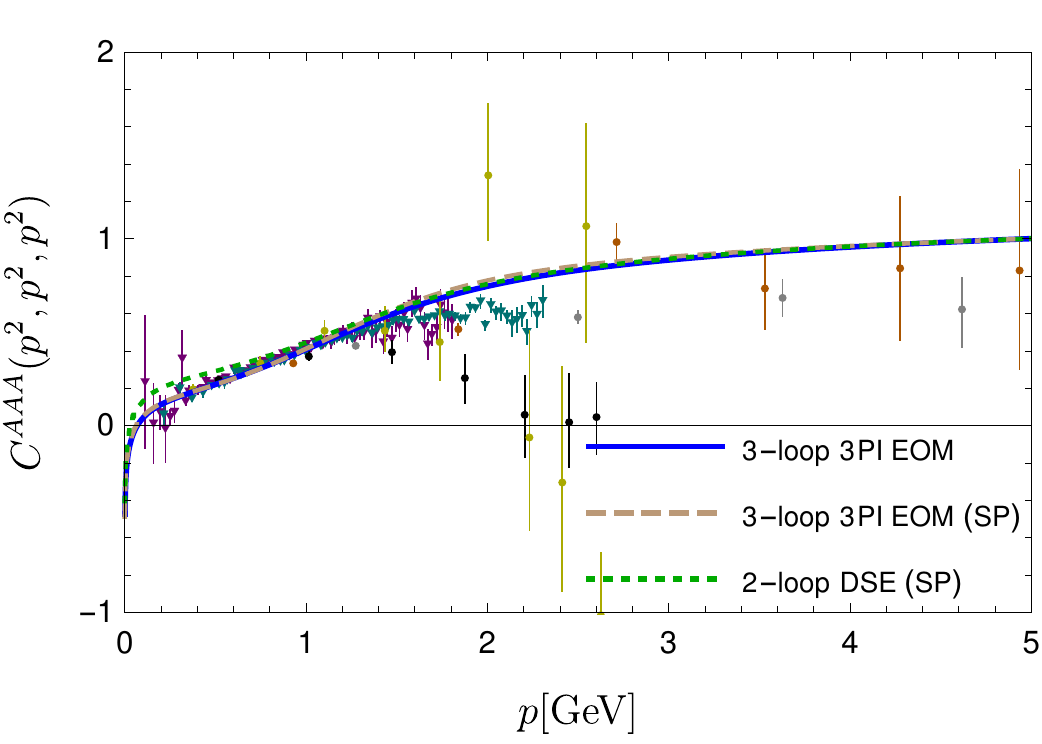}\hfill
	\includegraphics[width=0.48\textwidth]{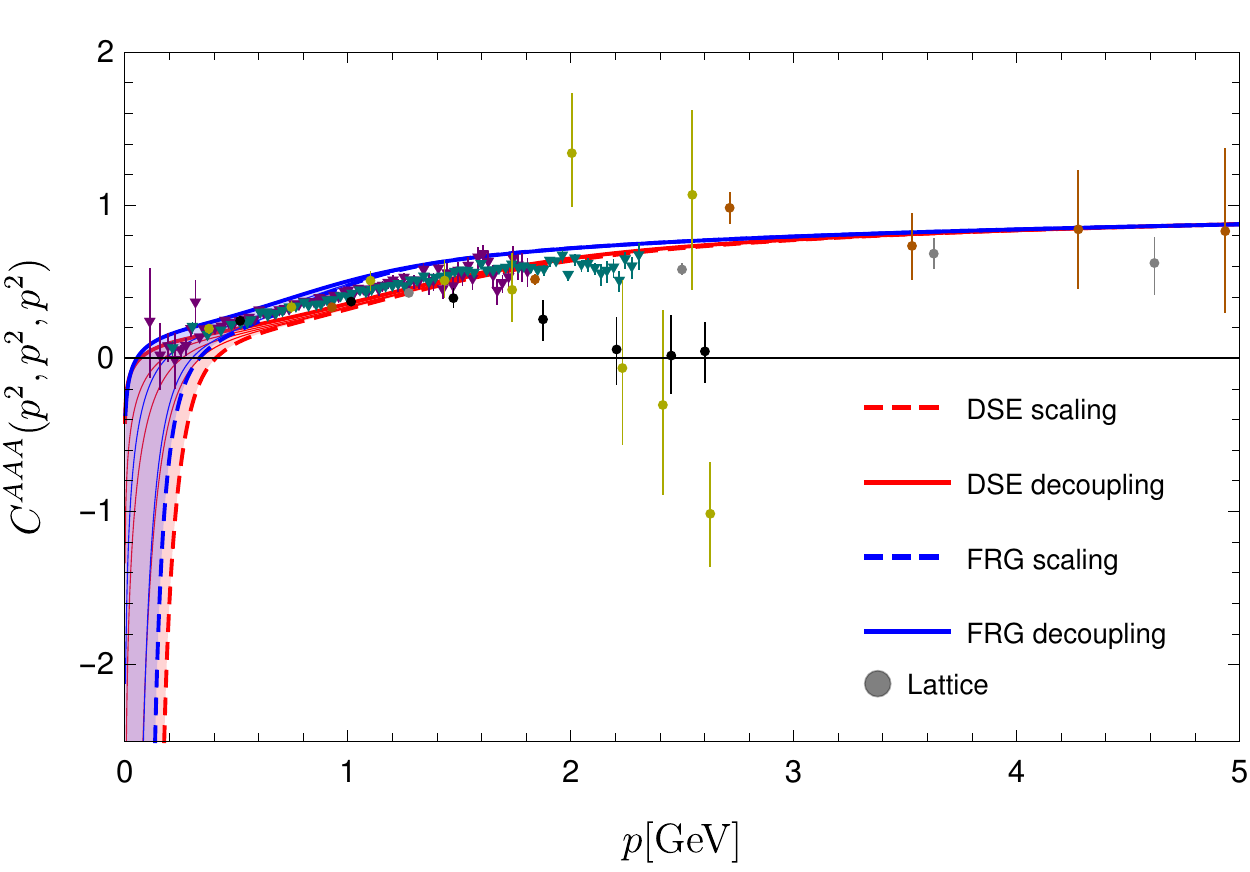}
	\caption{Left: Three-gluon vertex dressing function from the equation of motion of the 3PI effective action with full and restricted kinematics (SP) and from the two-loop truncated DSE.
	Right: Three-gluon vertex dressing function at the symmetric point in comparison to FRG \cite{Cyrol:2016tym}.
	Lattice data from \cite{Cucchieri:2008qm,Sternbeck:2017ntv}.
	}
	\label{fig:tg_comp}
\end{figure*}

The propagator and vertex results shown in \fref{fig:props} and \fref{fig:tg_comp}, respectively, compare favourably with lattice results \cite{Huber:2020keu}.
But what is even more remarkable is that several other tests have been performed that indicate the quantitative reliability of the results.
The most relevant is the fact that results from different functional equations agree.
This is illustrated in \fref{fig:tg_comp} where the 3PI results of the three-gluon vertex are compared with results from the functional renormalization group and from its Dyson-Schwinger equation (DSE).
Also shown is the negligible effect of restricting the three-gluon vertex to a single kinematic variable \cite{Huber:2020keu}.
This feature was already contained in early results of the vertex \cite{Blum:2014gna} but became only visible after switching to the proper variable \cite{Eichmann:2014xya} and was demonstrated also on the lattice \cite{Pinto-Gomez:2022brg}.
Other evidence of the reliability of the results is provided by the comparison of the couplings from the different vertices down to a few GeV \cite{Huber:2020keu} which could not be achieved with previous truncations, e.g., \cite{Blum:2014gna,Cyrol:2014kca}.

The effects of various quantities not taken into account by the employed truncation were also tested.
An enlarged tensor basis for the three-gluon vertex was studied in \cite{Eichmann:2014xya}.
The corresponding dressing functions are small compared to the leading one.
The effect of the four-gluon vertex was studied in three dimensions, where also only a small impact was found \cite{Huber:2016tvc}.
And finally, the two remaining four-point functions, the two-ghost-two-gluon and four-ghost vertices, were calculated in \cite{Huber:2017txg}.
Their inclusion in the calculation of other correlation functions again had only tiny or even vanishing effects \cite{Huber:2017txg}.
Thus, it seems that for the current level of precision all the important pieces are included and the obtained propagators and vertices provide a reliable input for the calculation of bound states.

\subsection{Truncation}
\label{sec:truncation}

% BSE & kernels
\begin{figure*}[tb]
	\includegraphics[width=0.65\textwidth]{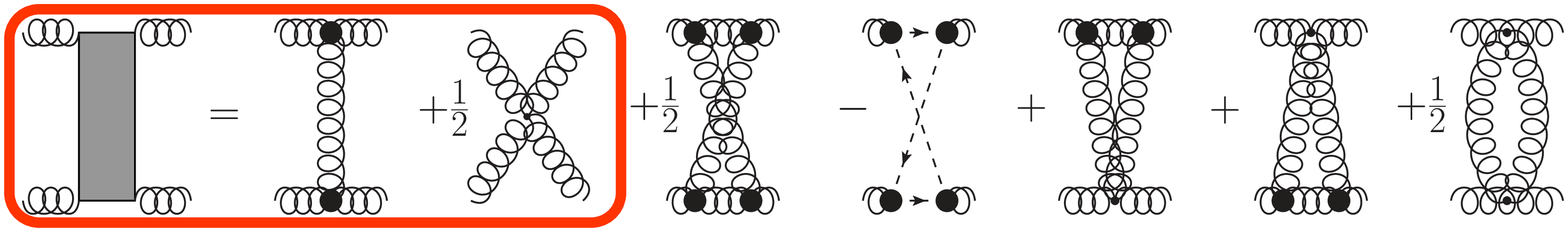}\\ 
	\vskip4mm
	\includegraphics[height=1.2cm]{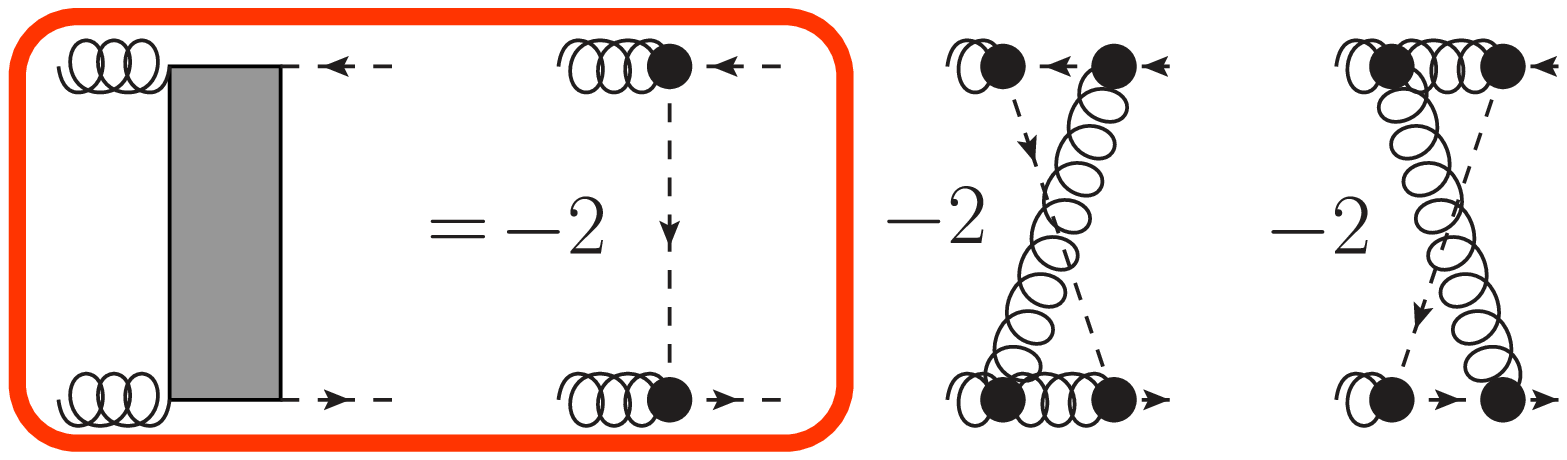}\hfill
	\includegraphics[height=1.2cm]{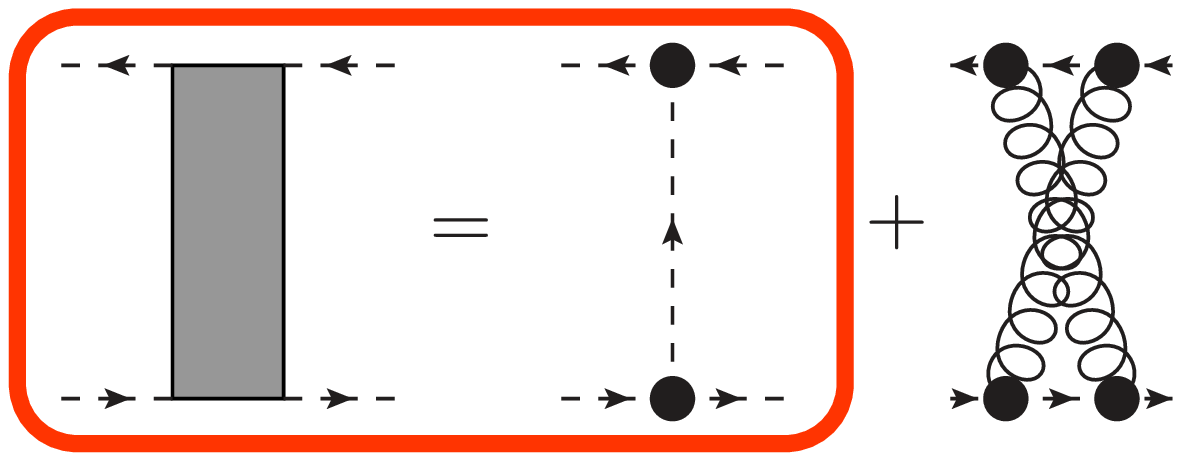}\hfill
	\includegraphics[height=1.2cm]{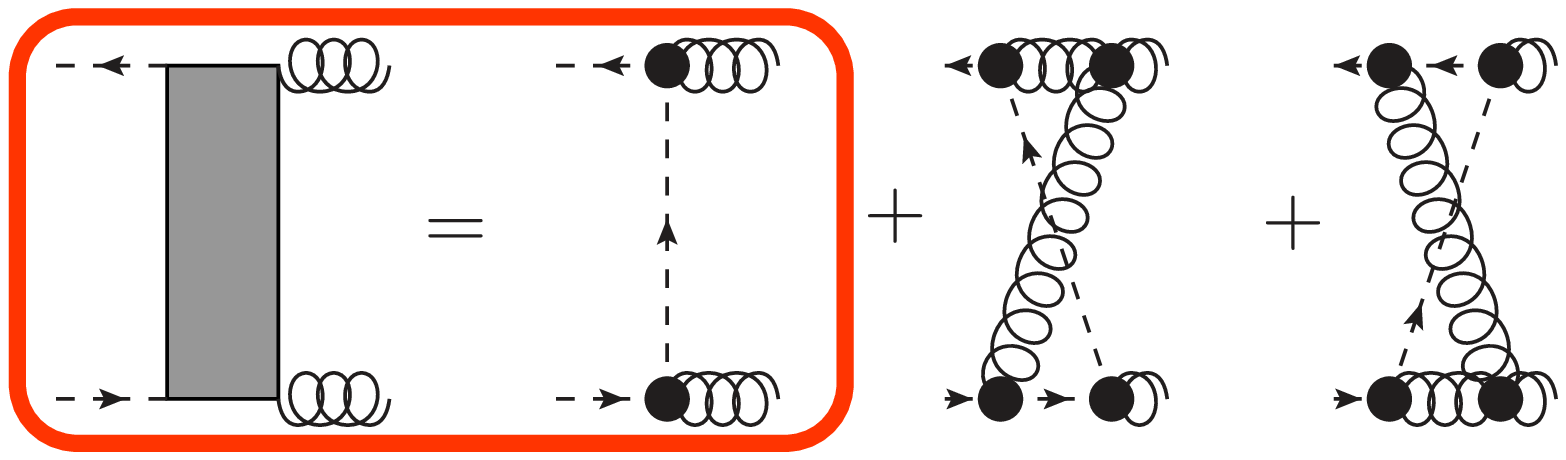}
	\caption{
		Interaction kernels from the three-loop 3PI effective action.
		All propagators are dressed; black disks represent dressed vertices.
		The red rectangles denote the main truncation.
		\label{fig:kernels}
	}
\end{figure*}

As the equations for the correlation functions, the kernels are derived from the 3PI effective action truncated at three loops \cite{Berges:2004pu,Sanchis-Alepuz:2015tha}.
This leads to the kernels shown in \fref{fig:kernels}.
As one can see, the kernels contain one-loop terms which lead to two-loop terms when inserted into the BSE.
Such diagrams are cumbersome to deal with and require substantially more computational power.
Initially, only the diagrams in the red rectangles were included in the calculation of the kernels \cite{Huber:2020ngt}.
However, to be fully self-consistent with the input, the two-loop diagrams were later added.
For the scalar glueball, a tiny effect (below 1\% for the ground state mass, 2\% for the first excited state) was found \cite{Huber:2022rhh} when the two-loop diagrams of the gluon-gluon kernel were included.
The two-loop diagrams of the other kernels are expected to have even less impact, because already the one-loop kernels are subleading.
For the pseudoscalar glueball the shift in the mass was barely visible \cite{Huber:2021zqk}.
In this case, all two-loop diagrams were included as the pseudoscalar glueball does not contain diagrams with ghosts.

\section{Conclusions}
\label{sec:conclusions}

The potential of functional equations in hadron physics ranges from answering fundamental questions about the nature of hadrons, see as an example Ref.~\cite{Eichmann:2020oqt} for the $\sigma$ meson, to providing quantitative results from first principles.
The crucial ingredient in the latter calculations is a high-quality input from a self-consistent and self-contained truncation of the equations of motion.
In this contribution we illustrated this with the example of the glueball spectrum for pure Yang-Mills theory.

\section*{Acknowledgments}

This work was supported by the DFG (German Research Foundation) grants FI 970/11-1 and FI 970/11-2 and by the BMBF under contract No. 05P21RGFP3.
This work has also been supported by Silicon Austria Labs (SAL), owned by the Republic of Austria, the Styrian Business Promotion Agency (SFG), the federal state of Carinthia, the Upper Austrian Research (UAR), and the Austrian Asso­ci­a­tion for the Elec­tric and Elec­tronics Industry (FEEI).

\bibliography{literature_glueballs_confxv}

\end{document}